\documentclass[10pt, conference]{IEEEtran}

\IEEEoverridecommandlockouts
\usepackage{cite}
\usepackage{amsmath,amssymb,amsfonts}
\usepackage{algorithmic}
\usepackage{graphicx}
\usepackage{textcomp}
\usepackage{xcolor}
\def\BibTeX{{\rm B\kern-.05em{\sc i\kern-.025em b}\kern-.08em
    T\kern-.1667em\lower.7ex\hbox{E}\kern-.125emX}}
\begin{document}

\title{Attentional networks for music generation}

\author{\IEEEauthorblockN{Gullapalli Keerti\IEEEauthorrefmark{1}, A N Vaishnavi\IEEEauthorrefmark{1}, Prerana Mukherjee\IEEEauthorrefmark{1}, A Sree Vidya\IEEEauthorrefmark{1},Gattineni Sai Sreenithya \IEEEauthorrefmark{1}, Deeksha Nayab\IEEEauthorrefmark{1}}
\IEEEauthorblockA{\textit{\IEEEauthorrefmark{1}Indian Institute of Information Technology, Sri City, Chittoor, Andhra Pradesh, India}}
 \{keerti.g17, vaishnavi.a17, prerana.m, sreevidya.a17, saisreenithya.g17, deeksha.n17\}@iiits.in }



\maketitle

\begin{abstract}
Realistic music generation has always remained as a challenging problem as it may lack structure or rationality. In this work, we propose a deep learning based music generation method in order to produce old style music particularly JAZZ with rehashed melodic structures utilizing a Bi-directional Long Short Term Memory (Bi-LSTM) Neural Network with Attention. Owing to the success in modelling long-term temporal dependencies in sequential data and its success in case of videos, Bi-LSTMs with attention serve as the natural choice and early utilization in music generation. We validate in our experiments that Bi-LSTMs with attention are able to preserve the richness and technical nuances of the music performed.





\end{abstract}


\section{Introduction}
\label{sec:intro}

Artistic skills made musicians use different modern computer tools to make their music more better and versatile. They can thus create a variety of expressive styles that are appealing.
For some imaginative purposes, gifted artists utilize conventional 
media or current PC apparatuses to make an assortment of expressive styles that are exceptionally engaging yet issue happens when they arrive at the bottleneck of making it more realistic. 
Fine subtleties is especially 
significant ingredient of music age. 
 Tuning in to fascinating music and if there is some approach to produce music naturally, especially good quality music at that point, it's a major jump in the realm of music industry.



Fortunately enough, neural networks applied to music had an alternate confidence during the AI winter in 1970s. During the period from 1988 to 2009,   a significant progression led to the traction in this field. It was pioneered by the work of Lewis and Todd \cite{todd1988sequential,lewis1988creation} in the 80s to the work of Eck and Schmidhuber \cite{eck2002finding} where we have traced a long way. Their work first utilized LSTMs in music generation. In \cite{johnson2017generating}, authors stated that LSTMs are able to capture the medium-scale melodic structure
in music pretty well. When trained on  sufficient audio data, they are also able to generate novel melodies.
Due to the recent success in speech synthesis models, particularly with WaveNet \cite{oord2016wavenet} raw audio files are increasingly used in music generation.


\begin{figure}[htbp]
\centerline{\includegraphics[width=0.5\textwidth]{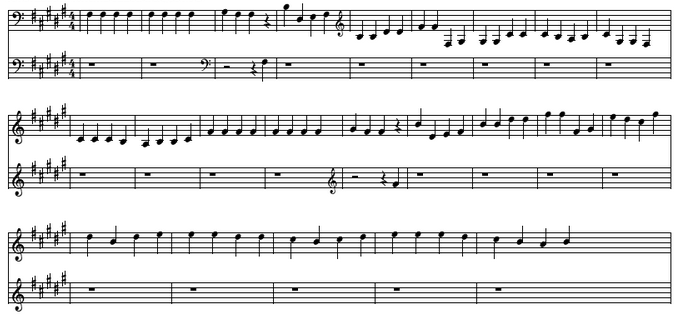}}
\caption{Sheet Music of the song: "The Last Farewell" by Roger Whittaker.}
\label{fig1}
\end{figure}

\begin{figure}[htbp]
\centerline{\includegraphics[width=0.5\textwidth]{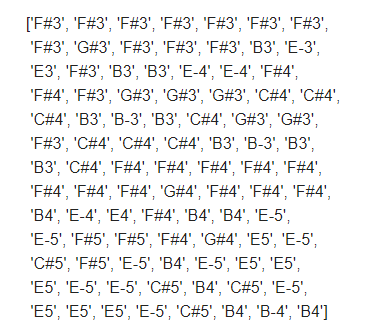}}
\caption{Musical Notes (Extracted from MIDI file) of the song: "The Last Farewell".}
\label{fig2}
\end{figure}

In this work, we utilize attention based bidirectional LSTM networks to produce old style Jazz music with rehashed melodic structures.The input to the network is the raw audio file in MIDI format. It is preprocessed and converted into the musical notes. This is then fed to a 512 layered Bi-LSTM network with attention which is again followed by an LSTM layer and flattened to generate the consecutive musical notes. This can be appended to the original MIDI file. Fig. \ref{fig1} shows the music sheet of the song ``The Last Farewell" in MIDI format. Fig. \ref{fig2} gives a list of musical notes in ASCII characters of the corresponding music file. Table \ref{tab1} shows the batch construction for the used dataset.


\begin{table}[]
\centering
\caption{Batch construction for the JAZZ ML ready MIDI dataset: Batch Size 64 Characters}
\tiny
\begin{tabular}{|l|l|l|l|l|l|}
\hline
                                                  & Batch-1                                                 & Batch-2                                                 & ...                                                     & Batch-150                                               & Batch-151                                               \\ \hline
0                                                 & 0...63                                                  & 64...127                                                & ...                                                     & 9536...9599                                             & 9600...9663                                             \\ \hline
1                                                 & 9701...9764                                             & 9765...9829                                             & ...                                                     & 19237...19300                                           & 19301...19364                                           \\ \hline
\begin{tabular}[c]{@{}l@{}}.\\ .\\ .\end{tabular} & \begin{tabular}[c]{@{}l@{}}...\\ ...\\ ...\end{tabular} & \begin{tabular}[c]{@{}l@{}}...\\ ...\\ ...\end{tabular} & \begin{tabular}[c]{@{}l@{}}...\\ ...\\ ...\end{tabular} & \begin{tabular}[c]{@{}l@{}}...\\ ...\\ ...\end{tabular} & \begin{tabular}[c]{@{}l@{}}...\\ ...\\ ...\end{tabular} \\ \hline
14                                                & 135814â€¦135877                                           & 135878â€¦135941                                           & ...                                                     & 145350â€¦145413                                           & 145414â€¦145477                                           \\ \hline
15                                                & 145515â€¦145578                                           & 145579â€¦145642                                           & ...                                                     & 155051â€¦155114                                           & 155115â€¦155178                                           \\ \hline
.                                                 & ...                                                     & ...                                                     & ...                                                     & ...                                                     & ...                                                     \\ \hline
\label{tab1}
\end{tabular}
\end{table}

\section{Background and Related Work}
\label{sec:background}
The pioneering work on profound learning based music is done  by Chen et al \cite{chen2001creating}, where the authors produce
a music with just a single tune. The authors perform  preprocessing steps in the data such as they removed speckled notes, rests, and 
off-tune harmonies. They addressed one of the principle issues which is the absence of structure in the generated music with machine learning based methods. In order to circumvent this issue, the two possible solutions are as follows: 
1. to construct music with melodic beat, increasingly complex structure, and using a wide range of notes counting specked notes, longer harmonies and rests. 
or
2. to construct a model equipped for adapting to long- term dependency structure and having the capacity to assemble new tune. 

Liu et al \cite{liu2014bach} also address a similar issue. They suggested that 
the music generated by their approach didn't appropriately recognize the song and different fragments of the 
piece and thus not able to capture the entire essence that most of the old style music pieces have. 
Eck et al.\cite{eck2002first} utilized two distinctive LSTM based networks i) to learn harmony structure and nearby note structure, ii) to realize longer dependency conditions so as to attempt to become familiar with a song and hold it all through 
the piece. This enabled the authors to generate music that never deviates a long way from the first harmony 
progression song. However, this method could just handle limited number of harmonies and cannot 
make a progressively assorted blend of notes. 
Boulanger-Lewandowski et al. \cite{boulanger2012modeling} attempted to manage the test of learning complex polyphonic structure in music. They utilized a Recurrent Temporal Restricted Boltzmann machine (RTRBM) so as to demonstrate unconstrained polyphonic music. Utilizing the RTRBM modelling enabled them to speak to a confounded dispersion over each time step as opposed to a solitary token as in 
most character language models. This model is also able to handle the issue of polyphony in the music generated.
Drewes et al. \cite{drewes2007algebra} proposed a strategy to utilize algebra to create music in a linguistic way with the 
help of tree-based models. Markov chains \cite{schulze2010music} and Markov hidden units can also be utilized to devise a numerical model to 
produce music. 

After the leap forward in AI, numerous new models and techniques were proposed in the field of music age. 
Depiction of different AI empowered procedures can be found in \cite{boulanger2012modeling,hadjeres2017interactive,browne2001system} including a probabilistic model utilizing RNNs, 
Anticipation RNN and Recursive Artificial Neural Networks (RANN), an adaptation of artificial neural networks 
\cite{abraham2005artificial} for creating the consequent notes, resulting note duration and rhythm. Generative Adversarial Networks 
(GANs) \cite{goodfellow2014generative} are also effectively utilized in generating melodic notes where the model consists of two networks, generator that is responsible for generating random information and discriminator that is responsible for assessing created arbitrary information for realness against the 
original data. MuseGAN \cite{dong2018musegan} is a generative adversarial network that creates representative multi-track music. Next, in the subsequent sections we provide the background on different components relevant to this work.

\subsection{RNN}
Recurrent Neural Networks (RNNs) include intermittent associations inside the hidden layers between past and current states in the NN. This capacity of memory storage makes it extremely helpful in applications such as discourse handling and music composition. The primary issue with a standard RNN is that it stores the data of just the previously attended state; this implies the setting expands just a single strand back. This isn't extremely helpful in music composition where the start of the tune may be quite significant than in the center and the end too.

\subsection {LSTM}

Long Short Term Memory networks  generally called "LSTMs" are an extraordinary sort of RNN, equipped for adapting long term conditions. They were presented by Hochreiter and Schmidhuber, and were refined and promoted by numerous individuals. They work colossally well on a huge assortment of issues, and are currently broadly utilized.

All intermittent neural networks have the type of a chain of rehashing modules of neural networks. In standard RNNs, this rehashing module will have an extremely basic structure, for example, a solitary tanh layer. 

Regularly used architecture of LSTM units have a cell and three regulators. Cell is the memory part of the LSTM unit.
Regulators comprises of input, output and forget gate. The dependencies between the subsequent input notes is taken care by the cell.
The sigmoid layer (activation function of LSTM) yields numbers somewhere in the range of zero and one, depicting the amount of every part ought to be let through. An estimation of zero signifies "let nothing through," while an estimation of one signifies "let everything through". 
\subsection{Attention based LSTM}
Attention is a later advancement that really takes care of our center issue.
 This enables to take care of specific segments of the contribution at some random moment and utilize those segments to help produce portions of the yield as opposed to simply the last output of the LSTM layer. 

 \begin{equation}
 c_{i} = \sum_{j=1}^n {\alpha_{ij}h_{j}} 
 \end{equation}
where $\alpha_{ij}$ are weights that define the consideration of hidden states in each output.
$c_{i}$ is the context vector for output $y_{i}$ which is the sum of hidden states of input sequence.
$h_{j}$ is the encoder network's hidden state. 
\begin{equation}
    \alpha_{ij} = align({y_{i}},{x_{j}})
\end{equation}

Here $\alpha_{ij}$ tells how well subsequent notes  $y_{i}$ and $x_{j}$ are aligned.   

\begin{equation}
   \alpha_{ij}=\frac{exp(score(s_{i-1}, h_j)))}{\sum_{j'=1}^nexp(scores_{i-1},h_j'))}
\end{equation}

The $\alpha_{ij}$ gives the softmax of predefined alignment score.
\begin{equation}
 score(s_i,h_j) = {v_a}^T tanh(W_a[s_i,h_j])   
\end{equation}
where ${v_{a}}\ and {W_{a}}$  are weight matrices that are learned in the alignment model.
${s_{i}}$ is the decoder network's hidden states.



\subsection{Bidirectional LSTM}
Bidirectional LSTMs are an expansion of conventional LSTMs that can improve model execution on arrangement order issues. In issues where all timesteps of the information arrangement are accessible, Bidirectional LSTMs train two rather than one LSTM on the information grouping.

\subsection{MIDI}

The Musical Instrument Digital Interface format (MIDI or .mid) is used to store message rules which contain note pitches, their volume, speed, start and end timestamp, phrases and so forth. It doesn't store songs like sound formats, however it stores data that is equipped for producing future melodic notes. These rules can be deciphered by a sound card which uses a wavetable (table of recorded sound waves) to make an understanding of the MIDI messages into genuine stable information. It very well may be deciphered by midi player studio, for example, Fruity Loops (FL) Studio or standard sequencers like Synthesia. Musical notation software like MuseScore or Finale can make a translation of midi into editable sheet music; this empowers customers to make music in regular music documentation on their PCs and they may listen to it by MIDI players.

\section{Proposed Methodology}
\begin{figure*}[htbp]
\centerline{\includegraphics[width=1.0\textwidth]{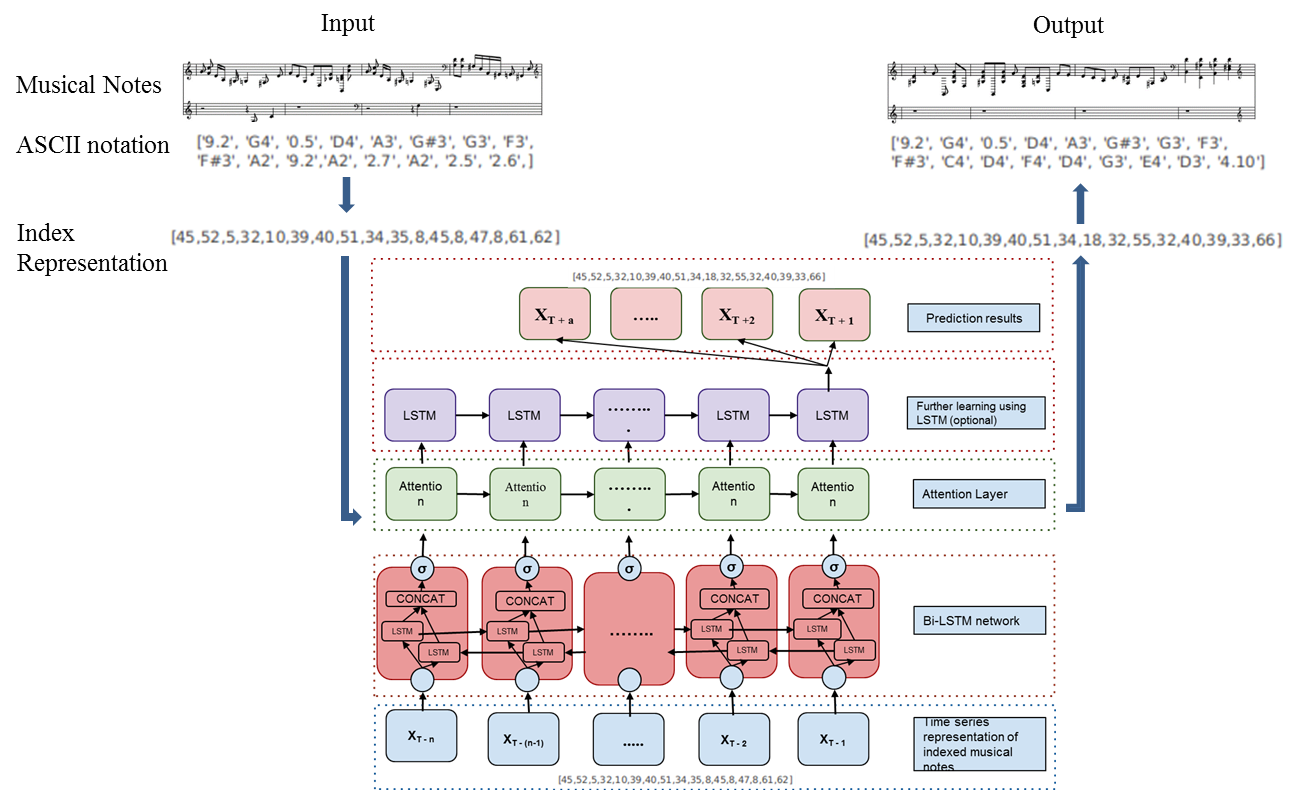}}
\caption{Architecture diagram}
\label{fig3}
\end{figure*}

Fig. \ref{fig3} outlines the proposed architecture for music generation.

\subsection{Preprocessing}
To prepare the model, the MIDI records should have to be  changed over into a structure that can be encoded as numeric information to feed the Bidirectional LSTM network. We have utilized a software package Music21 to preprocess the data. We included the rests and duration (Rhythm) as opposed to just having notes and chords. We utilized the Music21 library to take our MIDI records and convert them into a stream object which consists of rhythms, various voices, and notes/harmony/rest objects, with a related instrument and time duration. It parses all the MIDI files and annexed each note/harmony/rest-duration combination to a vector representation which is subdivided into 100-dimensional note samples. Those subdivided samples are then fed into the Bidirectional LSTM for training in order to generate the consecutive notes.

The data consists of two parts: i) Notes and ii) Chords. Pitch, octave, and offset of the Note are also covered as note objects. 
Pitch refers to the recurrence of the sound, or how high or low it is and is indicated with the letters [A, B, C, D, E, F, G], with A being the most elevated and G being the least. 
Octave refers to the set of pitches one may use on a piano. 
Offset refers to where the note is situated in the musical piece. 
Similarly, Chord objects are basically a placeholder for a lot of notes that are played simultaneously. 
Intuitively, we may observe that to generate music precisely our neural network should have the option to anticipate the upcoming note or harmony. Bi-LSTMs networks can mimic that properly. In our experimentation, the training set comprises of various notes and harmonies. 
The notes generally have shifting interims between them. We can have numerous notes with hardly a pause in between and afterward followed by a rest period where no note is played for a brief timeframe. 

\subsection{Music Generation using Attention based LSTM}

To achieve our objective of generating old style music with rehashed melodic structure, we have utilized Bi-LSTM network with attention layer.
We have utilized 1x100 dimensional input notes sample size into a Bidirectional LSTM with 512 cells, followed by an attention layer, subsequent LSTM layer with 512 nodes. 
Finally we have a 3400 node Dense layer with softmax predictions,
3400 denotes the number of possible unique note/chord/rest-duration combinations in the input data. 
 We have utilized dropout to reduce over-fitting issues. Further, we utilized categorical cross entropy loss and rmsprop as the optimizer function.
The second LSTM enables further learning these interdependencies between the notes and harmonies.

\section{Experimentation and Results}
\subsection{Dataset Discription }
We used Jazz ML ready MIDI dataset\footnote{https://www.kaggle.com/saikayala/jazz-ml-ready-midi} to train our model. The dataset comprises of 818 diverse Jazz music melodies.  There are 804 distinct notes which are annexed in a unique way i.e. each distinct character which is represented in ASCII is mapped to unique numerical value. The dataset comprises of:
\begin{itemize}
\item The list of notes extracted from the midi file,
\item Number of notes and 
\item List of unique notes for each midi file.
\end{itemize}

\subsection{Experimentation Results}

\begin{table}[]
\caption{Performance analysis on music generation.}
\scriptsize
\begin{tabular}{clll}
\hline
\textbf{Methods}                                                    & \multicolumn{1}{c}{\textbf{\begin{tabular}[c]{@{}c@{}}Categorical\\ Cross Entropy Loss\end{tabular}}} & \multicolumn{1}{c}{\textbf{RMSE}} & \multicolumn{1}{c}{\textbf{MSE}} \\ \hline \hline
\textbf{MidiNet \cite{yang2017midinet}}                                                    & 0.1203                                                                                                & 0.7112                            & 0.5058                           \\ 
\textbf{Adrien Ycart and E. Benetos \cite{ycart2017study}}                                & 0.2317                                                                                                & 0.9904                            & 0.9808                           \\ 
\textbf{LSTM}                                                       & 0.5097                                                                                                & 1.0919                            & 1.1924                           \\ 
\textbf{\begin{tabular}[c]{@{}c@{}}LSTM +\\ Attention\end{tabular}} & 0.2286                                                                                                & 0.8924                            & 0.7864                           \\ 
\textbf{Bi-LSTM + Attention + LSTM}                                 & 0.1069                                                                                                & 0.6694                            & 0.4481                           \\ \hline
\end{tabular}
\label{tab2}
\end{table}

\begin{figure}[htbp]
\centerline{\includegraphics[width=0.5\textwidth]{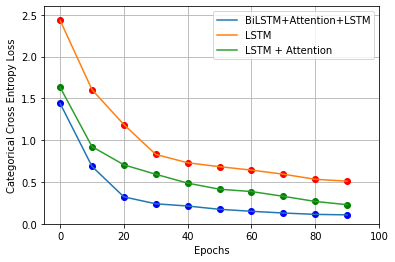}}
\caption{Graph: Categorical Cross Entropy Loss }
\label{fig4}
\end{figure}
\begin{figure*}[htbp]
\centerline{\includegraphics[width=1.0\textwidth]{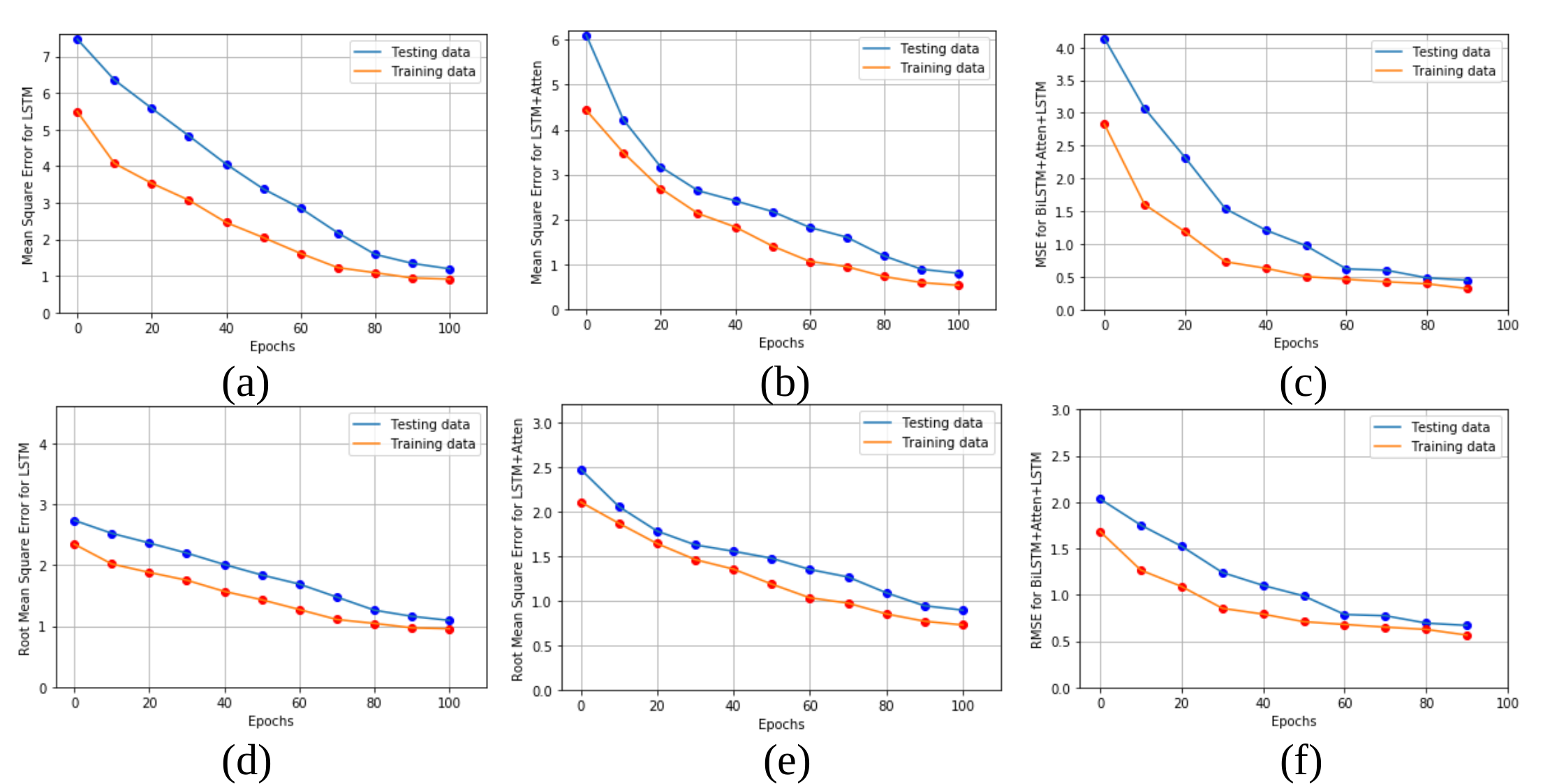}}
\caption{Performance Evaluation Graph: (a)-(c) shows Mean Square Error for LSTM, LSTM+attention and Bi-LSTM+attention respectively. (d)-(f) shows Root Mean Square Error for LSTM, LSTM+attention and Bi-LSTM+attention respectively. }
\label{fig5}
\end{figure*}

The output generated file is compared with the original file to find out the deviations from the input sequence. 
As seen in Table \ref{tab2}, the cross entropy loss and errors (Root mean square error (RMSE) and mean square error (MSE)) between the subsequent notes is compared for three variants: LSTM, LSTM with Attention, Bidirectional-LSTM with Attention and LSTM. LSTM performs the least than other two as it gives high error rates in all the three error functions. In order to improve the learning capability, attention layer is added to the LSTM and it is found that it perfoms better than vanilla LSTM network.
Finally, we observe that in all cases Bidirectional-LSTM with Attention and followed by stacked LSTM gives the best results. Fig. \ref{fig4} shows the categorical cross entropy loss for the three variants of LSTMs considered in this work. The learning curves are plotted in Fig. \ref{fig5}.

\subsection{Comparison with related works}
 In \cite{ycart2017study}, the author discusses about polyphonic midi sequences using LSTM networks. On comparing with the LSTM and LSTM+Attention it showed better results, but it showed a relatively high error rate when compared with Bi-directional LSTM with Attention and LSTM. Because music is all about leraning the patterns and inorder to recreate it we needed a model to understand these patterns and to be able generate subsequent notes to make a pleasant tune.Hence, attention plays a key role in our model. In \cite{yang2017midinet}, the error rate is relatively lower than \cite{ycart2017study} and when compared with Bidirectional LSTM with Attention and LSTM the error rates are almost similar. The differentiating factor between these two is how well the generated notes match with original one.






\section{Conclusions and Future Work}
In this paper, we presented bidirectional LSTM with Attention and LSTM with the objective of producing music that is coherent and good to hear. Our proposed model improved the structure of the generated music by understanding the patterns in it and training them until a better accuracy and minimal error rates is achieved. 
Given the ongoing trends in AI in music industry, we envisage that the current work presents progressively complex models and information portrayals that successfully captures the fundamental melodic structure.




\bibliographystyle{IEEEtran}
\bibliography{refs}

\begin{thebibliography}{10}
\providecommand{\url}[1]{#1}
\csname url@samestyle\endcsname
\providecommand{\newblock}{\relax}
\providecommand{\bibinfo}[2]{#2}
\providecommand{\BIBentrySTDinterwordspacing}{\spaceskip=0pt\relax}
\providecommand{\BIBentryALTinterwordstretchfactor}{4}
\providecommand{\BIBentryALTinterwordspacing}{\spaceskip=\fontdimen2\font plus
\BIBentryALTinterwordstretchfactor\fontdimen3\font minus
  \fontdimen4\font\relax}
\providecommand{\BIBforeignlanguage}[2]{{%
\expandafter\ifx\csname l@#1\endcsname\relax
\typeout{** WARNING: IEEEtran.bst: No hyphenation pattern has been}%
\typeout{** loaded for the language `#1'. Using the pattern for}%
\typeout{** the default language instead.}%
\else
\language=\csname l@#1\endcsname
\fi
#2}}
\providecommand{\BIBdecl}{\relax}
\BIBdecl

\bibitem{todd1988sequential}
P.~Todd, ``A sequential network design for musical applications,'' in
  \emph{Proceedings of the 1988 connectionist models summer school}, 1988, pp.
  76--84.

\bibitem{lewis1988creation}
J.~Lewis, ``Creation by refinement: A creativity paradigm for gradient descent
  learning networks,'' in \emph{International Conf. on Neural Networks}, 1988,
  pp. 229--233.

\bibitem{eck2002finding}
D.~Eck and J.~Schmidhuber, ``Finding temporal structure in music: Blues
  improvisation with lstm recurrent networks,'' in \emph{Proceedings of the
  12th IEEE workshop on neural networks for signal processing}.\hskip 1em plus
  0.5em minus 0.4em\relax IEEE, 2002, pp. 747--756.

\bibitem{johnson2017generating}
D.~D. Johnson, ``Generating polyphonic music using tied parallel networks,'' in
  \emph{International conference on evolutionary and biologically inspired
  music and art}.\hskip 1em plus 0.5em minus 0.4em\relax Springer, 2017, pp.
  128--143.

\bibitem{oord2016wavenet}
A.~v.~d. Oord, S.~Dieleman, H.~Zen, K.~Simonyan, O.~Vinyals, A.~Graves,
  N.~Kalchbrenner, A.~Senior, and K.~Kavukcuoglu, ``Wavenet: A generative model
  for raw audio,'' \emph{arXiv preprint arXiv:1609.03499}, 2016.

\bibitem{chen2001creating}
C.-C. Chen and R.~Miikkulainen, ``Creating melodies with evolving recurrent
  neural networks,'' in \emph{IJCNN'01. International Joint Conference on
  Neural Networks. Proceedings (Cat. No. 01CH37222)}, vol.~3.\hskip 1em plus
  0.5em minus 0.4em\relax IEEE, 2001, pp. 2241--2246.

\bibitem{liu2014bach}
I.~Liu, B.~Ramakrishnan \emph{et~al.}, ``Bach in 2014: Music composition with
  recurrent neural network,'' \emph{arXiv preprint arXiv:1412.3191}, 2014.

\bibitem{eck2002first}
D.~Eck and J.~Schmidhuber, ``A first look at music composition using lstm
  recurrent neural networks,'' \emph{Istituto Dalle Molle Di Studi Sull
  Intelligenza Artificiale}, vol. 103, p.~48, 2002.

\bibitem{boulanger2012modeling}
N.~Boulanger-Lewandowski, Y.~Bengio, and P.~Vincent, ``Modeling temporal
  dependencies in high-dimensional sequences: Application to polyphonic music
  generation and transcription,'' \emph{arXiv preprint arXiv:1206.6392}, 2012.

\bibitem{drewes2007algebra}
F.~Drewes and J.~H{\"o}gberg, ``An algebra for tree-based music generation,''
  in \emph{International Conference on Algebraic Informatics}.\hskip 1em plus
  0.5em minus 0.4em\relax Springer, 2007, pp. 172--188.

\bibitem{schulze2010music}
W.~Schulze and B.~Van Der~Merwe, ``Music generation with markov models,''
  \emph{IEEE MultiMedia}, no.~3, pp. 78--85, 2010.

\bibitem{hadjeres2017interactive}
G.~Hadjeres and F.~Nielsen, ``Interactive music generation with positional
  constraints using anticipation-rnns,'' \emph{arXiv preprint
  arXiv:1709.06404}, 2017.

\bibitem{browne2001system}
C.~B. Browne, ``System and method for automatic music generation using a neural
  network architecture,'' Oct.~2 2001, uS Patent 6,297,439.

\bibitem{abraham2005artificial}
A.~Abraham, ``Artificial neural networks,'' \emph{Handbook of measuring system
  design}, 2005.

\bibitem{goodfellow2014generative}
I.~Goodfellow, J.~Pouget-Abadie, M.~Mirza, B.~Xu, D.~Warde-Farley, S.~Ozair,
  A.~Courville, and Y.~Bengio, ``Generative adversarial nets,'' in
  \emph{Advances in neural information processing systems}, 2014, pp.
  2672--2680.

\bibitem{dong2018musegan}
H.-W. Dong, W.-Y. Hsiao, L.-C. Yang, and Y.-H. Yang, ``Musegan: Multi-track
  sequential generative adversarial networks for symbolic music generation and
  accompaniment,'' in \emph{Thirty-Second AAAI Conference on Artificial
  Intelligence}, 2018.

\bibitem{yang2017midinet}
L.-C. Yang, S.-Y. Chou, and Y.-H. Yang, ``Midinet: A convolutional generative
  adversarial network for symbolic-domain music generation,'' \emph{arXiv
  preprint arXiv:1703.10847}, 2017.

\bibitem{ycart2017study}
A.~Ycart, E.~Benetos \emph{et~al.}, ``A study on lstm networks for polyphonic
  music sequence modelling.''\hskip 1em plus 0.5em minus 0.4em\relax ISMIR,
  2017.

\end{thebibliography}

\end{document}